\documentstyle[amstex,amssymb,epsfig,12pt]{article}

\input{tcilatex}

\begin{document}

\begin{titlepage}
\bigskip \begin{flushright}
WATPPHYS-TH01/07
\end{flushright}


\vspace{1cm}

\begin{center}
{\Large \bf{Vortex Holography}}\\
\end{center}
\vspace{2cm}
\begin{center}
M.H. Dehghani\footnote{%
EMail: hossein@@avatar.uwaterloo.ca; On leave from Physics
Dept., College of Sciences, Shiraz University, Shiraz, Iran},  A.M. Ghezelbash\footnote{
EMail: amasoud@@avatar.uwaterloo.ca; On leave from Department of Physics,
Az-zahra University, Tehran, Iran} and R. B. Mann{%
\footnote{%
EMail: mann@@avatar.uwaterloo.ca}} \\
Department of Physics, University of Waterloo, \\
Waterloo, Ontario N2L 3G1, CANADA\\
\vspace{2cm}
\vspace{2cm}
\today\\
\end{center}

\begin{abstract}
We show that the Abelian Higgs field equations in the four dimensional
anti de Sitter spacetime have a vortex line solution.
This solution, which has cylindrical symmetry in  AdS$_4$, is a generalization of the flat 
spacetime Nielsen-Olesen string. We show that the vortex induces a deficit angle
in the AdS$_4$ spacetime that is proportional to  its mass density.
Using the AdS/CFT correspondence, we show that the mass  density of the string 
is uniform and dual to the discontinuity of a logarithmic derivative of 
correlation function of the boundary scalar  operator.

\end{abstract}
\end{titlepage}\onecolumn

\begin{center}
\bigskip 
\end{center}

\section{Introduction}

\bigskip

The general idea behind the holographic conjecture is that a conformal field
theory defined on a $d$-dimensional boundary of a $(d+1)$-dimensional
spacetime (with prescribed asymptotic behaviour) provides a sufficient
description of the physics of quantum gravity in that spacetime.\ Its most
concrete manifestation has been due to the work of \ Maldacena \cite{Mal}
and Witten \cite{W} concerning the relation of \ large ${\cal N}$ gauge
theories and conformal field theories. Specifically, the conjectured
correspondence is between the large ${\cal N}$ \ limit of superconformal
gauge theories and supergravity on $AdS_{d+1}$ spaces \cite{Mal}, one which
has also been studied in connection with non-extremal black-hole physics 
\cite{Sfet}.

\bigskip More precisely, consider the partition function of any field theory
on $AdS_{d+1}$ defined by 
\begin{equation}
Z_{AdS}[\phi _{0}]=\int_{{\phi _{o}}}{\cal D}\phi \text{ }e^{-S(\phi )}\text{
}  \label{PAR}
\end{equation}
where $\phi _{0}$ is the finite field defined on the boundary of $AdS_{d+1}$
and the integration is over the field configurations $\phi $ that approach $%
\phi _{0}$ when one goes from the bulk of $AdS_{d+1}$ to its boundary. The
conjectured correspondence states that $Z_{AdS}$ is identified with the
generating functional $Z_{CFT}$ of the boundary conformal field theory given
by 
\begin{equation}
Z_{CFT}[\phi _{0}]=<\exp \left( \int_{\partial {\cal M}{_{d}}}d^{d}x\sqrt{g}%
{\cal O}\phi _{0}\right) >  \label{PAR2}
\end{equation}
for a quasi-primary conformal operator ${\cal O}$ on the boundary $\partial 
{\cal M}{_{d}}$ of $AdS_{d+1}$ \cite{F,W,G}. This correspondence has been
explicated for a free massive scalar field and a free $U(1)$ gauge theory 
\cite{W}; other examples, such as interacting massive scalar \cite{Mu1},
free massive spinor \cite{Hen} and interacting scalar-spinor fields \cite
{Kav} have also been investigated, along with classical gravity and type-IIB
string theory \cite{Lu, Banks, Chal}. In all these cases, the exact
partition function (\ref{PAR}) is given by the exponential of the action
evaluated for a classical field configuration which solves the classical
equations of motion, and explicit calculations show that the evaluated
partition function is equal to the generating functional (\ref{PAR2}) of
some conformal field theory with a quasi-primary operator of a certain
conformal weight.

These results encourage the expectation that an understanding of quantum
gravity in a given spacetime (at least one that is asymptotically AdS) can
be carried out by studying \ instead its dual theory, defined on the
boundary of spacetime at infinity. \ This general ``holographic principle''
-- of which the AdS/CFT correspondence can be viewed as a special case --
has among other things the advantage that there are significantly fewer
degrees of freedom in the holographic dual theory than there are in the bulk
theory.

Consequently, it is natural to inquire about the relationship between
objects in the bulk and their dual holographic counterparts, and much effort
has been expended in this direction. \ For example the asymptotic behaviour
of bulk fields is directly related to one-point functions in the CFT \cite
{bala3}, and it has been shown that the radial position of a position of a
source particle following a bulk geodesic is encoded in the size and shape
of an expectation-value bubble in the CFT \cite{ulf}. \ When distinct bulk
solutions have the same asymptotic behaviour it has been argued that
non-local objects in the CFT are required to distinguish them; specifically,
propagator kinks in Green's functions in the CFT can be used to detect the
presence of point particles in $(2+1)$-dimensional spacetimes that are
asymptotically AdS \cite{Bala}.

\bigskip

In this paper we extend these considerations to extended objects in the
bulk. Specifically, we consider topologically non-trivial solutions of
Abelian-Higgs field equations in the four dimensional anti de Sitter space.
Such vortex solutions have long been known in flat space \cite{NO}, and have
been of some interest in black hole physics in recent years since they
provide a specific example of stable hair for the Schwarzchild black hole in 
$(3+1)$ dimensions\thinspace \cite{Achu}. In this paper we investigate how a
gauge vortex can be holographically represented via the AdS/CFT
correspondence. These objects have some features in common with those of
lower-dimensional point-particles: a cross-sectional slice at some fixed
polar angle yields a spacetime approximately equivalent to that of a $(2+1)$%
-dimensional asymptotically AdS spacetime with a point-particle. However
unlike the point particle the vortex solution extends all the way to the
boundary, entailing different considerations of \ its relationship to the
CFT.

The model we consider will appear as the low energy limit of any string
theory containing the minimal supersymmetric standard model in some
(low-energy) limit, along with a mechanism in which supersymmetry is broken
only by super-renormalizable terms. The Higgs field potential is given by
the linear combination of the D-term of a scalar superfield potential, the
F-term of another scalar field potential, and the most general
superrenormalizable supersymmetry-breaking term. This term yields a
potential and gauge couplings of the form we consider \cite{Weinberg}. \ The
Abelian Higgs model has also recently been shown to be equivalent to the
theory of a massive Kalb-Ramond string interacting with the worldsheet of
the vortex in the limit of large Higgs coupling and thin core radius \cite
{Koma}.

The outline of our paper is as follows. In section two, we present a
solution of \ the $U(1)$ Abelian Higgs equations with non-zero winding
number $(3+1)$-dimensional anti de Sitter spacetime (AdS$_{4}$), and justify
that these solutions describe a vortex-line structure. We then show in
section three that this solution induces (for thin strings and to leading
order in the gravitational coupling) a deficit angle in AdS$_{4}$. Since
this solution has the same local asymptotic behaviour as pure AdS$_{4}$, we
therefore expect \cite{Bala}\ that a non-local quantity in the boundary CFT
will be needed for its holographic description. In section 4 we show this to
be the case: we consider the two point correlation function of the dual
boundary conformal scalar operator and show that there exists a kink in this
correlation function which encodes the mass per unit length of the vortex.
We compute this in section 5 using the boundary counterterm approach, and
find the mass density to be uniform. A concluding section rounds out the
paper.

\section{Abelian Higgs Vortex in AdS$_{4}$}

We take the Abelian Higgs Lagrangian in AdS$_{4}$ as follows,

\begin{equation}
{\cal L}(\Phi ,A_{\mu })=\frac{1}{2}({\cal D}_{\mu }\Phi )^{\dagger }{\cal D}%
^{\mu }\Phi -\frac{1}{16\pi }{\cal F}_{\mu \nu }{\cal F}^{\mu \nu }-\xi
(\Phi ^{\dagger }\Phi -\eta ^{2})^{2}  \label{Lag}
\end{equation}
where $\Phi $ is a complex \ scalar Klein-Gordon field, ${\cal F}_{\mu \nu }$
is the field strength of \ the electromagnetic field $A_{\mu }$ and ${\cal D}%
_{\mu }=\nabla _{\mu }+ieA_{\mu }$ in which $\nabla _{\mu }$ is the
covariant derivative. We employ Planck units $G=\hbar =c=1$ which implies
that the Planck mass is equal to one, and write the AdS$_{4}$ spacetime
metric in the form 
\begin{equation}
ds^{2}=-(1+\frac{r^{2}}{l^{2}})dt^{2}+\frac{1}{(1+\frac{r^{2}}{l^{2}})}%
dr^{2}+r^{2}(d\theta ^{2}+\sin ^{2}\theta \,d\varphi ^{2})  \label{adsmetr1}
\end{equation}
\qquad

Defining the real fields $X(x^{\mu }),\omega (x^{\mu }),P_{\mu }(x^{\nu })$
by the following equations

\begin{equation}
\begin{tabular}{l}
$\Phi (x^{\mu })=\eta X(x^{\mu })e^{i\omega (x^{\mu })}$ \\ 
$A_{\mu }(x^{\nu })=\frac{1}{e}(P_{\mu }(x^{\nu })-\nabla _{\mu }\omega
(x^{\mu }))$%
\end{tabular}
\label{XPomegadef}
\end{equation}
and employing a suitable choice of gauge, one could rewrite the Lagrangian (%
\ref{Lag}) and the equations of motion in terms of these fields as: 
\begin{equation}
{\cal L(}X,P_{\mu })=-\frac{\eta ^{2}}{2}(\nabla _{\mu }X\,\nabla ^{\mu
}X+X^{2}P_{\mu }P^{\mu })-\frac{1}{16\pi e^{2}}F_{\mu \nu }F^{\mu \nu }-\xi
\eta ^{4}(X^{2}-1)^{2}  \label{Lag2}
\end{equation}

\begin{equation}
\begin{tabular}{l}
$\nabla _{\mu }\nabla ^{\mu }X-XP_{\mu }P^{\mu }-4\xi \eta ^{2}X(X^{2}-1)=0$
\\ 
$\nabla _{\mu }F^{\mu \nu }+4\pi e^{2}\eta ^{2}P^{\nu }X^{2}=0$%
\end{tabular}
\label{eqmo2}
\end{equation}
where $F^{\mu \nu }=\nabla ^{\mu }P^{\nu }-\nabla ^{\nu }P^{\mu }$ is the
field strength of the corresponding gauge field $P^{\mu }$. Note that the
real field $\omega $ is not itself a physical quantity. Superficially it
appears not to contain any physical information. However if $\omega $ is not
single valued this is no longer the case, and the resultant solutions are
referred to as vortex solutions \cite{NO}. \ In this case the requirement
that $\Phi $ field be single-valued implies that the line integral of $%
\omega $ over any closed loop is $\pm 2\pi n$ where $n$ is an integer. In
this case the flux of electromagnetic field $\Phi _{H\text{ \ }}$passing
through such a closed loop is quantized with quanta $2\pi /e.$

We seek a vortex solution for the Abelian Higgs Lagrangian (\ref{Lag2}) in
the background of AdS$_{4}$.This solution can be interpreted as a string
like object in the background metric (\ref{adsmetr1}). We consider the
static cylindrically symmetric case with the gauge choice, 
\begin{equation}
P^{\mu }(r,\theta )=(0;0,0,P(r,\theta ))  \label{Pgauge}
\end{equation}
where $\mu $ goes from $0$ to $3$, corresponding to the coordinates $%
t,r,\theta ,\varphi $ in the metric (\ref{adsmetr1}). The equations of
motion (\ref{eqmo2}) are 
\begin{equation}
\begin{tabular}{l}
$(1+\frac{r^{2}}{l^{2}})\frac{\partial ^{2}P(r,\theta )}{\partial r^{2}}+%
\frac{2}{r}(2+3\frac{r^{2}}{l^{2}})\frac{\partial P(r,\theta )}{\partial r}+%
\frac{1}{r^{2}}\frac{\partial ^{2}P(r,\theta )}{\partial \theta ^{2}}+3\frac{%
\cot \theta }{r^{2}}\frac{\partial P(r,\theta )}{\partial \theta }+\frac{6}{%
l^{2}}P(r,\theta )$ \\ 
$-4\pi e\eta ^{2}(eP(r,\theta )+\frac{\csc ^{2}\theta }{r^{2}}%
)X^{2}(r,\theta )=0$%
\end{tabular}
\label{eqp}
\end{equation}

\begin{equation}
\begin{tabular}{l}
$(1+\frac{r^{2}}{l^{2}})\frac{\partial ^{2}X(r,\theta )}{\partial r^{2}}+2(%
\frac{1}{r}+\frac{2r}{l^{2}})\frac{\partial X(r,\theta )}{\partial r}+\frac{1%
}{r^{2}}\frac{\partial ^{2}X(r,\theta )}{\partial \theta ^{2}}+\frac{1}{r^{2}%
}\frac{\partial X(r,\theta )}{\partial \theta }\cot \theta -4\xi \eta
^{2}X^{3}(r,\theta )$ \\ 
$-\{\frac{1}{r^{2}}\csc ^{2}\theta +e^{2}r^{2}\sin ^{2}\theta P^{2}(r,\theta
)+2eP(r,\theta )-4\xi \eta ^{2}\}X(r,\theta )=0$%
\end{tabular}
\label{eqxx}
\end{equation}
We seek a cylindrically symmetric solution, one for which 
\begin{equation}
P(r,\theta )=-A(\rho )/\rho ,\text{ \ }X(r,\theta )=X(\rho )
\label{PXcylsym}
\end{equation}
where $\rho =r\sin \theta $. \ We thus obtain the following equations of
motion 
\begin{equation}
(1+\frac{\rho ^{2}}{l^{2}})\frac{d^{2}A}{d\rho ^{2}}+\frac{dA}{d\rho }(\frac{%
1}{\rho }+\frac{4\rho }{l^{2}})+A(\frac{2}{l^{2}}-\frac{1}{\rho ^{2}})+4\pi
e\eta ^{2}(eA-\frac{1}{\rho })X^{2}=0  \label{eqa}
\end{equation}
\begin{equation}
(1+\frac{\rho ^{2}}{l^{2}})\frac{d^{2}X}{d\rho ^{2}}+(\frac{1}{\rho }+\frac{%
4\rho }{l^{2}})\frac{dX}{d\rho }-4\xi \eta ^{2}X^{3}-\{(\frac{1}{\rho }%
-eA)^{2}-4\xi \eta ^{2}\}X=0  \label{eqx}
\end{equation}
As expected, in the limit $l\rightarrow \infty $, the equations (\ref{eqa})
and (\ref{eqx}) reduce to those whose solutions describe the well known
Nielsen-Olesen vortex in flat spacetime. In this case the solution for the
gauge field is represented by a combination of Bessel functions which at
large distance decay exponentially.

Exact analytical solutions to the above equations (\ref{eqa}) and (\ref{eqx}%
) are not known. However if we assume that $X$ \ becomes constant at large $%
\rho $, $\ X_{0}=X(\rho \rightarrow \infty )=1$, which is the necessary
condition to have a vortex line solution, then we are able to analytically
solve (\ref{eqa}) for the gauge field, obtaining 
\begin{equation}
A(\rho )=\frac{1}{e\rho }+S(\rho )\{c_{1}+c_{2}\int^{\rho }\frac{d\zeta }{%
\zeta (\zeta ^{2}+l^{2})^{3/2}S^{2}(\zeta )}\}  \label{Asoln}
\end{equation}
where the function $S(\rho )$ is given by

\begin{equation}
S(\rho )=\rho \text{ }_{\text{ }2}F_{\text{ }1}(\frac{5}{4}+\frac{1}{4}\sqrt{%
1+16\pi e^{2}\eta ^{2}l^{2}},\frac{5}{4}-\frac{1}{4}\sqrt{1+16\pi e^{2}\eta
^{2}l^{2}},2;-\frac{\rho ^{2}}{l^{2}})  \label{A(R)}
\end{equation}
and$\ _{\text{ }2}F_{\text{ }1}(a,b,c;x)$ \ is the usual hypergeometric
function. \ To obtain the behaviour of solution (\ref{Asoln}) in the limit $%
l\rightarrow \infty $, one may use the following relation between the
hypergeometric function and the Bessel function, 
\begin{equation}
\lim\Sb \varkappa \rightarrow \infty  \\ \varrho \rightarrow \infty  \endSb 
\text{ }_{2}F_{1}(\varkappa ,\varrho ,\nu +1;-\frac{x^{2}}{4\varkappa
\varrho })=\frac{2^{\nu }\Gamma (\nu +1)J_{\nu }(x)}{x^{\nu }}
\label{golden}
\end{equation}
in which $\varkappa $ and $\varrho $ could go to infinity through real or
complex values \cite{Abro}.

The solution (\ref{Asoln}) for the gauge potential then reduces to a
combination of the Bessel functions $J_{1}(\rho )$ and $N_{1}(\rho )$. By a
suitable choice of constants of integration $c_{1}$ and $c_{2}$ we obtain 
\begin{equation}
A(\rho )=\frac{1}{e\rho }-\frac{\pi }{2e}c{\cal H}_{1}^{(1)}(i\sqrt{4\pi
e^{2}}\eta \rho )  \label{Ainf}
\end{equation}
where $c$ is a constant and ${\cal H}_{1}^{(1)}(i\sqrt{4\pi e^{2}}\eta \rho
) $ is the Hankel function of order one, which in the large $\rho $ limit
has the well behaved decaying exponential form analogous to that observed in 
\cite{NO}.

\begin{figure}[tbp]
\begin{center}
\epsfig{file=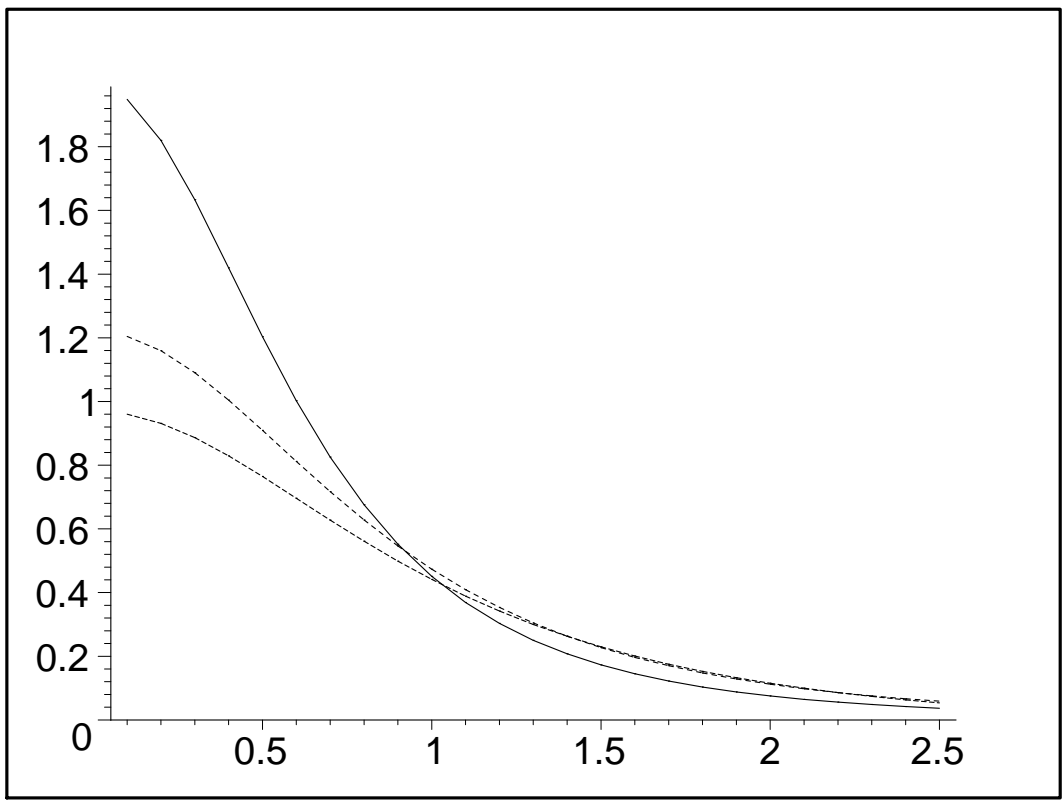,width=0.8\linewidth}
\end{center}
\caption{$H(\protect\rho )$ for $l=1$ (solid), $l=2$ (dotted), $l=5$
(dashed).}
\label{Fig1}
\end{figure}

The magnitude of the magnetic field ${\bf H}(\rho )$, which is given by 
\begin{equation}
H(\rho )=\frac{1}{2\pi \rho }\frac{d\Phi _{H}}{d\rho }=\frac{1}{\rho }\frac{%
d(\rho A(\rho ))}{d\rho }  \label{Hfieldrho}
\end{equation}
is 
\begin{eqnarray}
H(\rho ) &=&\frac{2}{\rho }\{S(\rho )-\frac{\rho ^{3}}{2l^{2}}(\frac{3}{2}%
-\pi e\eta ^{2}l^{2})\text{ }_{2}F_{\text{ }1}(\frac{9}{4}+\frac{1}{4}\sqrt{%
1+16\pi e^{2}\eta ^{2}l^{2}},\frac{9}{4}-\frac{1}{4}\sqrt{1+16\pi e^{2}\eta
^{2}l^{2}},3;-\frac{\rho ^{2}}{l^{2}})\}  \nonumber \\
&&\times \frac{A(\rho )-\frac{1}{e\rho }}{S(\rho )}+\frac{c_{2}}{\rho (\rho
^{2}+l^{2})^{3/2}S(\rho )}  \label{Hfield}
\end{eqnarray}
Again by using eq. (\ref{golden}) one could show that as $l$ goes to
infinity, the above solution reduces to 
\begin{equation}
H(\rho )=-i\frac{\pi }{2}c\eta {\cal H}_{0}^{(1)}(i\sqrt{4\pi e^{2}}\eta
\rho )  \label{Hiinf}
\end{equation}
where ${\cal H}_{0}^{(1)}(i\sqrt{4\pi e^{2}}\eta \rho )$ is the Hankel
function of order zero. The above relation (\ref{Hiinf}) is the same as that
of the magnetic field obtained for the string in the flat spacetime. In this
case in the large $\rho $ limit the magnetic field (\ref{Hiinf}) is
approximately 
\begin{equation}
H(\rho \rightarrow \infty )\sim \frac{1}{\sqrt{\rho }}e^{-\sqrt{4\pi e^{2}}%
\eta \rho }  \label{Hinf}
\end{equation}
Consider next the behaviour of the magnetic field $H(\rho )$ given by
equation (\ref{Hfield}) as a function of the distance from the string. For
values of $\ \eta <1$, and different values of \ the cosmological constant, $%
H(\rho )$ goes to zero very rapidly as $\rho $ goes to infinity, as
illustrated in Fig. (\ref{Fig1}). The characteristic length is defined to be
a distance from the string axis which measures the region in spacetime over
which the magnitude of $H(\rho )$ is appreciably different from zero. From
Fig. (\ref{Fig1}), we see that the characteristic length does not depend on
the cosmological constant and therefore one could consider the
characteristic length to be the same as the case of \ large $l$, which from
equation (\ref{Hinf}) is equal to $\lambda _{H}\sim \frac{1}{\sqrt{4\pi e^{2}%
}\eta }.$

Next we study the behaviour of the magnitude of the scalar field $X(\rho ).$
Eq.(\ref{eqx}) is approximately satisfied if 
\begin{equation}
X=X_{0}\simeq 1  \label{EqX0}
\end{equation}
where $X_{0}$ is the minimum of the potential in (\ref{Lag}). This minimum
is just the vacuum value of the field configuration. Denoting fluctuations
about this vacuum value by $\psi (\rho )$

\begin{equation}
X(\rho )=X_{0}+\psi (\rho )
\end{equation}
and expanding the potential in the Lagrangian about $X_{0},$ we have from
eq. (\ref{eqx}), 
\begin{equation}
\frac{\rho ^{2}}{l^{2}}\frac{d^{2}\psi }{d\rho ^{2}}+\frac{4\rho }{l^{2}}%
\frac{d\psi }{d\rho }-4\xi \eta ^{4}\psi (\rho )=0  \label{flucinf}
\end{equation}
where in deriving (\ref{flucinf}) from (\ref{eqx}), we have neglected terms
of order unity in the coefficients of the first and second terms involving
derivatives of the $X$ field with respect to the terms involving $\frac{\rho
^{2}}{l^{2}}$.

\begin{figure}[tbp]
\begin{center}
\epsfig{file=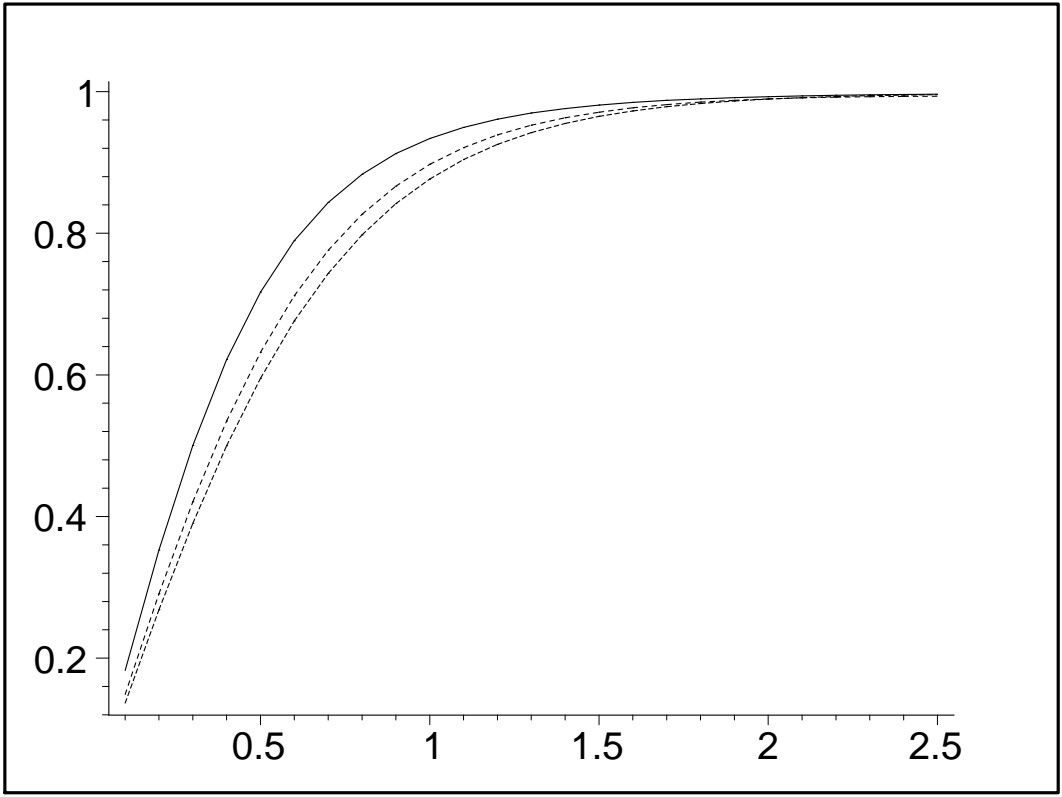,width=0.8\linewidth}
\end{center}
\caption{$X(\protect\rho )$ for $X_{0}=1$ and $l=1$ (solid), $l=2$ (dotted), 
$l=5$ (dashed).}
\label{Fig2}
\end{figure}

From (\ref{flucinf}) the approximate solution to eq. (\ref{eqx}) for large $%
\rho $ is 
\begin{equation}
X(\rho )\simeq X_{0}\{1-(\frac{\rho }{\rho _{0}})^{-\left( 3+\eta ^{2}\sqrt{%
9+16\xi l^{2}}\right) /2}\}  \label{Xinf}
\end{equation}
Figure (\ref{Fig2}) illustrates the behaviour of $X(\rho )$ for different
values of $l$ which is obtained by solving the field equations numerically.

As with the magnetic field, the $X$ \ field is nearly equal to its vacuum
value $X_{0}\sim 1$ everywhere except within a certain region $\rho \lesssim
\lambda _{X}$, which defines a characteristic length for this field. \ A
simple calculation shows that $X\simeq 1-\varepsilon $ for $\rho \simeq \rho
_{0}\left( 1+2\varepsilon /(3+\eta ^{2}\sqrt{9+16\xi l^{2}})\right) $ and so
the characteristic length $\lambda _{X}$ \ is of the order of\ $\rho _{0}$.
From \ figure (\ref{Fig2}) , we see that the value of $\lambda _{X}$ \ is
nearly independent of $\ l.$ \ It is easily seen that if the order of \
magnitude of $\lambda _{X}$ and $\lambda _{H}$ are nearly equal to each
other, then one has a well defined vortex line, or string. So the vacuum
state is described by $H=0$ and $X=X_{0}$, and the extension of the string
is given by $\lambda _{H}\sim \lambda _{X}.$ \ As one can see from figure (%
\ref{Fig2}), the value of$\ X$ \ is nearly equal to $X_{0}$ for \ $\rho $
greater than some $\rho $ which has the same order as $\rho _{0}=1$. This
shows that we can have a vortex solution for the field equations.

\section{Effect of the Vortex on AdS$_{4}$}

We now consider the effect of \ the vortex on the AdS$_{4}$ spacetime. This
entails finding the solutions of the coupled Einstein-Abelian Higgs
differential equations in AdS$_{4}$. This is a formidable problem even for
flat spacetime, and no exact solutions have been found \ for the flat
spacetime yet.

However\thinspace we can obtain some physical results by making judicious
approximations. First, we assume that the thickness of string is much
smaller that all the other relevant length scales. Second, we assume that
the gravitational effects of the string are weak enough so that the
linearized Einstein-Abelian Higgs differential equations are applicable.

\bigskip

For convenience, in this section we use the following form of the metric of
AdS$_{4}.$

\begin{equation}
ds^{2}=\exp (\frac{2z}{l})\left( -\exp (A(\widehat{\rho },z))d\widehat{t}%
^{2}+d\widehat{\rho }^{2}+\widehat{\rho }^{2}\exp (B(\widehat{\rho }%
,z))d\phi ^{2}\right) +\exp (C(\widehat{\rho },z))dz^{2}  \label{adsmetrgen}
\end{equation}
In the absence of the vortex, we must have $A(\widehat{\rho },z)=B(\widehat{%
\rho },z)=C(\widehat{\rho },z)=0$ , yielding 
\begin{equation}
ds^{2}=\exp (\frac{2z}{l})\left( d\widehat{\rho }^{2}+\widehat{\rho }%
^{2}d\phi ^{2}-d\widehat{t}^{2}\right) +dz^{2}  \label{adsmetr3}
\end{equation}
which is the metric for pure AdS$_{4}$. The transformation relations between
two metrics (\ref{adsmetr1}) and (\ref{adsmetr3}) are 
\begin{eqnarray}
&&  \nonumber \\
\widehat{\rho }\exp (\frac{z}{l}) &=&r\sin \theta  \label{trans} \\
\exp (\frac{z}{l}) &=&\frac{r}{l}\cos \theta +\sqrt{1+\frac{r^{2}}{l^{2}}}%
\cos (\frac{t}{l})  \label{trans2} \\
\widehat{t}\exp (\frac{z}{l}) &=&l\sqrt{1+\frac{r^{2}}{l^{2}}}\sin (\frac{t}{%
l})  \label{trans3}
\end{eqnarray}

Using the transformation $\rho =\widehat{\rho }\exp (\frac{z}{l})$ (which in
fact is equal to $\rho =r\sin \theta $) it is straightforward to show that
the Abelian Higgs equations in the background metric (\ref{adsmetr3}) are
simply equations (\ref{eqa} , \ref{eqx}). Employing the two assumptions
concerning the thickness of the vortex core and its weak gravitational
field, we solve the Einstein field equations $G_{\mu \nu }-\frac{3}{l^{2}}%
g_{\mu \nu }=-8\pi G{\cal T}_{\mu \nu },$ to this order of approximation by
taking $g_{\mu \nu }\simeq g_{\mu \nu }^{(0)}+g_{\mu \nu }^{(1)}$ , where $%
g_{\mu \nu }^{(0)}$ is given by (\ref{adsmetr3}), $g_{\mu \nu }^{(1)}$
includes the corrections which induce non-zero $A(\widehat{\rho },z),$ $B(%
\widehat{\rho },z)$ and $C(\widehat{\rho },z)$ in (\ref{adsmetrgen}) by
taking the energy-momentum tensor ${\cal T}_{\mu \nu }$ to be that
associated with the solution (\ref{Asoln}) and (\ref{Xinf}). \ Hence

\begin{equation}
G_{\mu \nu }^{(1)}-\frac{3}{l^{2}}g_{\mu \nu }^{(1)}=-8\pi {\cal T}_{\mu \nu
}^{(0)}  \label{Eineq}
\end{equation}
where ${\cal T}_{\mu \nu }^{(0)}$ is the energy momentum tensor of string
field in AdS$_{4\text{ }}$background metric (\ref{adsmetr3}), and $G_{\mu
\nu }^{(1)}$ is the correction to the Einstein tensor due to $g_{\mu \nu
}^{(1)}$.

After scaling the coordinate $\rho \rightarrow \frac{\rho }{\sqrt{\xi }\eta }%
,$ $z\rightarrow \frac{z}{\sqrt{\xi }\eta }$ the gauge field $P\rightarrow
P\xi \eta ^{2},$ and $l\rightarrow \frac{l}{\sqrt{\xi }\eta },$ then the
components of the energy momentum tensor of string in the background of AdS$%
_{4},$ $T_{\mu \nu }^{(0)}=\frac{{\cal T}_{\mu \nu }^{(0)}}{\xi \eta ^{4}}$
are given by, 
\begin{equation}
\begin{tabular}{l}
$T_{\widehat{t}}^{\widehat{t}(0)}(\rho )=(1+\frac{\rho ^{2}}{l^{2}})\{-\frac{%
1}{2}(\frac{dX}{d\rho })^{2}-2\beta \lbrack \frac{1}{4}\rho ^{2}(\frac{dP}{%
d\rho })^{2}+\rho P\frac{dP}{d\rho }+P^{2}]\}-\frac{1}{2}\rho
^{2}P^{2}X^{2}-(X^{2}-1)^{2}$ \\ 
$T_{\widehat{\rho }}^{\widehat{\rho }(0)}(\rho )=(1-\frac{\rho ^{2}}{l^{2}}%
)\{\frac{1}{2}(\frac{dX}{d\rho })^{2}+2\beta \lbrack \frac{1}{4}\rho ^{2}(%
\frac{dP}{d\rho })^{2}+\rho P\frac{dP}{d\rho }+P^{2}]\}-\frac{1}{2}\rho
^{2}P^{2}X^{2}-(X^{2}-1)^{2}$ \\ 
$T_{\varphi }^{\varphi (0)}(\rho )=(1+\frac{\rho ^{2}}{l^{2}})\{-\frac{1}{2}(%
\frac{dX}{d\rho })^{2}+2\beta \lbrack \frac{1}{4}\rho ^{2}(\frac{dP}{d\rho }%
)^{2}+\rho P\frac{dP}{d\rho }+P^{2}]\}+\frac{1}{2}\rho
^{2}P^{2}X^{2}-(X^{2}-1)^{2}$ \\ 
$T_{z}^{z(0)}(\rho )=(1-\frac{\rho ^{2}}{l^{2}})\{-\frac{1}{2}(\frac{dX}{%
d\rho })^{2}-2\beta \lbrack \frac{1}{4}\rho ^{2}(\frac{dP}{d\rho })^{2}+\rho
P\frac{dP}{d\rho }+P^{2}]\}-\frac{1}{2}\rho ^{2}P^{2}X^{2}-(X^{2}-1)^{2}$ \\ 
$\widetilde{T}_{z}^{\widehat{\rho }(0)}(\rho )=\frac{\rho }{l}\{(\frac{dX}{%
d\rho })^{2}+4\beta \lbrack \frac{1}{4}\rho ^{2}(\frac{dP}{d\rho })^{2}+\rho
P\frac{dP}{d\rho }+P^{2}]\}$%
\end{tabular}
\label{stresscyl}
\end{equation}
where $X$ and $P$ are the solutions of \ the string fields (\ref{Xinf}) and (%
\ref{Asoln}), $\beta =\xi /4\pi e^{2}$ is the Bogomolnyi parameter \cite{Bog}
and $\widetilde{T}_{z}^{\widehat{\rho }(0)}=T_{z}^{\widehat{\rho }(0)}\exp (%
\frac{z}{l})$.

\bigskip

\ As it is well known , the most general form of the metric of a
cylindrically symmetric spacetime has three arbitrary functions \cite{Syn}.
Since all diagonal components of $T_{\mu }^{\nu (0)}$ and $\widetilde{T}%
_{z}^{\widehat{\rho }(0)}$depend only on the combination $\rho =\widehat{%
\rho }\exp (\frac{z}{l})$, we assume the same for the functions $A(\widehat{%
\rho },z)$, $B(\widehat{\rho },z)$ and $C(\widehat{\rho },z)$ in the metric (%
\ref{adsmetrgen}) as well. Furthermore, we expect that these three functions
rapidly approach constants for large $\rho $\ , since the string
energy-momentum tensor falls off rapidly to zero outside the core of the
string. \ It is straightforward to show from the Einstein equations that $A(%
\widehat{\rho },z)=A(\rho )$ and $B(\widehat{\rho },z)=B(\rho )$ \ approach
non-zero constants whereas $C(\widehat{\rho },z)\rightarrow 0\ $in this
limit. The constant limit of $A(\rho )$ can be absorbed by a rescaling of
the time coordinate, whereas the Einstein equations imply for $B(\rho )$
that 
\begin{equation}
\frac{1}{2}(1+\frac{\rho ^{2}}{l^{2}})\frac{d^{2}B}{d\rho ^{2}}+\frac{1}{4}(%
\frac{dB}{d\rho })^{2}(1+\frac{\rho ^{2}}{l^{2}})+\frac{1}{\rho }\frac{dB}{%
d\rho }(1+\frac{2\rho ^{2}}{l^{2}})=-\varepsilon T_{\widehat{t}}^{\widehat{t}%
(0)}  \label{ein1}
\end{equation}
where $\varepsilon =8\pi G\eta ^{2}$. Introducing 
\begin{equation}
F(\rho )=\rho \exp (B(\rho )/2)  \label{cof}
\end{equation}
equation (\ref{ein1}) becomes 
\begin{equation}
\frac{2}{l^{2}}-\frac{1}{F}\frac{d}{d\rho }\{(1+\frac{\rho ^{2}}{l^{2}})%
\frac{dF}{d\rho }\}=\varepsilon T_{\widehat{t}}^{\widehat{t}(0)}
\label{eqfirst}
\end{equation}
which be integrated to find $F$ in terms of the mass per unit length of the
vortex 
\begin{equation}
\mu =\int_{0}^{2\pi }\int_{0}^{\rho _{0}}{\cal T}_{\widehat{t}}^{\widehat{t}%
(0)}\sqrt{^{(2)}g}d\widehat{\rho }d\varphi  \label{mpul}
\end{equation}
where $^{(2)}g$ is the determinant of two-dimensional metric induced on the
hypersurface $(\widehat{t},z)=(\widehat{t}_{0},z_{0})$ in the spacetime (\ref
{adsmetrgen}), \ and $\rho _{0\text{ }}$is the orthogonal distance from the
string. For $\rho \geq \rho _{0}$\ \ the energy-momentum tensor rapidly goes
to zero, and so $\mu $ becomes a non-zero constant. Using the relation (\ref
{eqfirst}) with the boundary conditions $F(0)=1,$ $F^{\prime }(0)=1$, \ we
obtain 
\begin{equation}
\mu =\frac{\gamma }{4\pi }(1+\frac{\rho _{0}^{2}}{l^{2}})+\frac{1}{2l^{2}}%
\int_{0}^{\rho _{0}}F(\rho )d\rho -\frac{1}{4}\frac{\rho _{0}^{2}}{l^{2}}
\label{massdef}
\end{equation}
where $2\gamma $ is the deficit angle 
\begin{equation}
\gamma =\pi (1-F^{\prime }(\rho _{0}))  \label{deficit}
\end{equation}
and so we see that the presence of the vortex induces a deficit angle in the
spacetime.

Note that in the special case of \ $F(\rho )=\rho $, there is no deficit
angle; from equations (\ref{massdef}) we must have $\mu =0.$ \ In the large-$%
l$ limit (\ref{massdef}) gives the correct known result between the vortex
mass density and deficit angle in flat spacetime \cite{Achu}. \ Numerical
integration of the remaining Einstein equations confirms the above ansatz
for the asymptotic behaviors of the functions in (\ref{adsmetr3}). A more
complete treatment of the vortex self-gravity in AdS$_{4\text{ }}$will be
dealt with elsewhere.

We therefore see that a thin vortex will, to leading order in the
gravitational coupling, yield the metric (\ref{adsmetr3}) but with a deficit
angle $2\gamma $ given by (\ref{massdef}). Using eqs. (\ref{trans}-\ref
{trans3}) this metric becomes (\ref{adsmetr1}), but now with the same
deficit angle in the $\phi $ coordinate. \ We shall henceforth take this to
be the metric induced by the vortex.

\section{Holographic Detection of \ a Stationary Vortex}

The scalar field $\Phi (t,R,\theta ,\phi )$ in the Lagrangian (\ref{Lag}) in
the AdS$_{4\text{ }}$spacetime\ is dual to the conformal operator ${\cal O}%
(t,\theta ,\phi )$ on the boundary of AdS$_{4}$ with the conformal weight $%
\Delta =3/2+\sqrt{9/4+\eta }$. The two-point correlation function of the
operator ${\cal O}$ in two distinct points $X$ and $Y$ on the boundary of AdS%
$_{4}$ according to the AdS/CFT correspondence is, 
\begin{equation}
<{\cal O(}X){\cal O}(Y)>=\frac{{\cal A}}{\mid X-Y\mid ^{2\Delta }}
\label{ccf}
\end{equation}
where ${\cal A}$ is a constant. This result is obtained \cite{W} from
regulating the boundary at infinity ($R/l=1$) by taking $R_{\varepsilon
}(t,\theta ,\phi )/l=1-\varepsilon (t,\theta ,\phi )$ and then letting $%
\varepsilon \rightarrow 0$, where $\varepsilon (t,\theta ,\phi )$ is a
smooth function on the boundary.

One can prove, following reasoning similar to that of ref. \cite{Bala}, that
the conformal two-point correlation function (\ref{ccf}) can be obtained by
evaluating the bulk propagator of a scalar field of mass $\sqrt{\eta }$
between the points $(R_{\varepsilon }(X),X)$ and $(R_{\varepsilon }(Y),Y)$
in the bulk of AdS$_{4}$ (up to a term which depends on the regulator $%
\varepsilon (t,\theta ,\phi )$). The bulk propagator of a scalar field is
given by 
\begin{equation}
G(x^{\mu },y^{\nu })=\int {\cal D}{\frak P\ }e^{i\Delta {\cal L}({\frak P})}
\label{bulkprop}
\end{equation}
where the integral is over all paths ${\frak P}$ between the points $x^{\mu
} $ and $\ y^{\nu }$and ${\cal L}({\frak P})$ is the proper geodesic length
of the path. Later, we will use the saddle point approximation to write the
right hand side of (\ref{bulkprop}), as the exponential of unique geodesic
length between the boundary points given by (\ref{len}). Although there is a
deficit angle $2\gamma $ due to the presence of the vortex given by the
equation (\ref{deficit}), we can use the saddle point approximation since
the spacetime is everywhere static. Furthermore the spacetime is locally
pure AdS$_{4}$ without any black hole structure, and so no issues of
causality arise, in contrast to the black hole situation which is considered
in ref. \cite{Lou}.

The AdS/CFT correspondence conjecture leads us to expect that some physical
information from the bulk space is encoded in the conformal correlation
functions. Once these are known, the natural question is how then to obtain
the corresponding bulk information (e.g. the mass of bulk fields, etc.) from
these correlation functions. To this end we shall evaluate the two-point
correlation function (\ref{ccf}). \ To proceed, we need to know the
structure of the geodesics of AdS$_{4}$, which we discuss in the next
subsection.

\subsection{Global AdS$_{4}$}

To find the structure of the geodesics of AdS$_{4}$, it is more convenient
to change the coordinate $r$ in (\ref{adsmetr1}) to $\chi $ 
\begin{equation}
r=l\sinh \chi  \label{rchi}
\end{equation}
for which the metric of the global AdS$_{4}$ becomes 
\begin{equation}
\begin{tabular}{l}
$ds^{2}=-\cosh ^{2}\chi dt^{2}+l^{2}d\chi ^{2}+l^{2}\sinh ^{2}\chi (d\theta
^{2}+\sin ^{2}\theta d\varphi ^{2})$ \\ 
$\ \ \ \ =-\frac{(1+R^{2})^{2}}{(1-R^{2})^{2}}dt^{2}+\frac{4l^{2}}{%
(1-R^{2})^{2}}\{dR^{2}+R^{2}(d\theta ^{2}+\sin ^{2}\theta d\varphi ^{2})\}$%
\end{tabular}
\label{ads4met}
\end{equation}
which is the Poincar\'{e} ball representation. The coordinate $R=\tanh (%
\frac{\chi }{2})$ ranges over the interval $[0,1]$. $\varphi $ has period $%
2\pi $, $\theta $ is between $0$ to $\pi $ and $t$ runs between $-\infty $
to $+\infty .$

Each constant time hypersurface in (\ref{ads4met}) is a Poincar\'{e} ball,
the boundary of which is a two-dimensional sphere.

\subsection{Geodesics}

\ Eliminating proper time from the geodesic equations yields the following
differential equations 
\begin{equation}
\frac{d^{2}\varphi (\theta )}{d\theta ^{2}}+\sin \theta \cos \theta (\frac{%
d\varphi (\theta )}{d\theta })^{3}+2\cot \theta \frac{d\varphi (\theta )}{%
d\theta }=0  \label{geoeq1}
\end{equation}
\begin{equation}
\begin{tabular}{l}
$\frac{d^{2}\chi (\theta )}{d\theta ^{2}}-2(\frac{d\chi (\theta )}{d\theta }%
)^{2}\coth (\chi (\theta ))+\sin \theta \cos \theta \frac{d\chi (\theta )}{%
d\theta }(\frac{d\varphi (\theta )}{d\theta })^{2}$ \\ 
$\ \ \ \ -\text{\ }\cosh (\chi (\theta ))\sinh (\chi (\theta ))\{1+\sin
^{2}\theta (\frac{d\varphi (\theta )}{d\theta })^{2}\}=0$%
\end{tabular}
\label{geoeq2}
\end{equation}
describing the path of minimal length between two arbitrary points on the
boundary of Poincar\'{e} ball.

To find the solutions of the eqs. (\ref{geoeq1}) and (\ref{geoeq2}), we note
that equal time geodesics of AdS$_{4\text{ }}$are circle segments which are
perpendicular to the three dimensional Poincar\'{e} ball parametrized by $%
R,\theta ,\varphi $ at $R=l$.

The different fields of the Abelian-Higgs theory have cylindrical symmetry
and so do not depend on the coordinate $z$. To evaluate quantities such as
the kink in the propagator (\ref{ccf}), it is therefore convenient to
consider two points on the boundary with the same $\theta $, thereby
respecting the cylindrical symmetry of the solution.

To find the geodesic path between two points on the boundary at the same $%
\theta =\theta _{0}$, which we denote by $M=(1,\theta _{0},-\varphi _{m})$
and $N=(1,\theta _{0},+\varphi _{m})$, we take a plane passing through both
of these points and the origin $O$. The geodesics are segments of circles in
this plane that are orthogonal to the intersection circle of the plane with
2-sphere boundary of the Poincar\'{e} ball. The angle $\theta $ between any
arbitrary vector in this plane and the $z$ axis is then a function of the
azimuthal angle $\varphi ,$ which can be written as 
\begin{equation}
\varphi =\varphi _{0}+\arccos (b\cot \theta )  \label{geo1}
\end{equation}
where $\varphi _{0}$ is an arbitrary constant and $b=\tan \theta
_{0}^{\prime }$, which $\theta _{0}^{\prime }$ is the minimum angle between
the $z$ axis and the plane \bigskip $OMN.$

We note that eq. (\ref{geo1}) is in fact the solution of (\ref{geoeq1}). To
find the solution of (\ref{geoeq2}), we observe that in the slanted plane $%
OMN$, we have 
\begin{equation}
\tanh \chi =\frac{a^{\prime }}{\cos (\varphi ^{\prime }-\varphi _{0}^{\prime
})}
\end{equation}
where $\varphi ^{\prime }$ is the azimuthal coordinate in the plane $OMN$, $%
\ a^{\prime }=\cos (\varphi _{m}^{\prime })$ in which $\pm \varphi
_{m}^{\prime }$ is the corresponding angles of \ the points $M$ and $N$ and $%
\varphi _{0}^{\prime }$ is a constant. Using the well known relation between
the coordinates $\varphi $ and $\varphi ^{\prime }$, one can get the
following equation 
\begin{equation}
\tanh \chi \cos \theta =\cos \theta _{0}  \label{geo2}
\end{equation}
We note that (\ref{geo2}) \ and (\ref{geo1}) satisfy the eq. (\ref{geoeq2})
and hence these two equations describe the equal time geodesic path. 
\begin{figure}[tbp]
\begin{center}
\epsfig{file=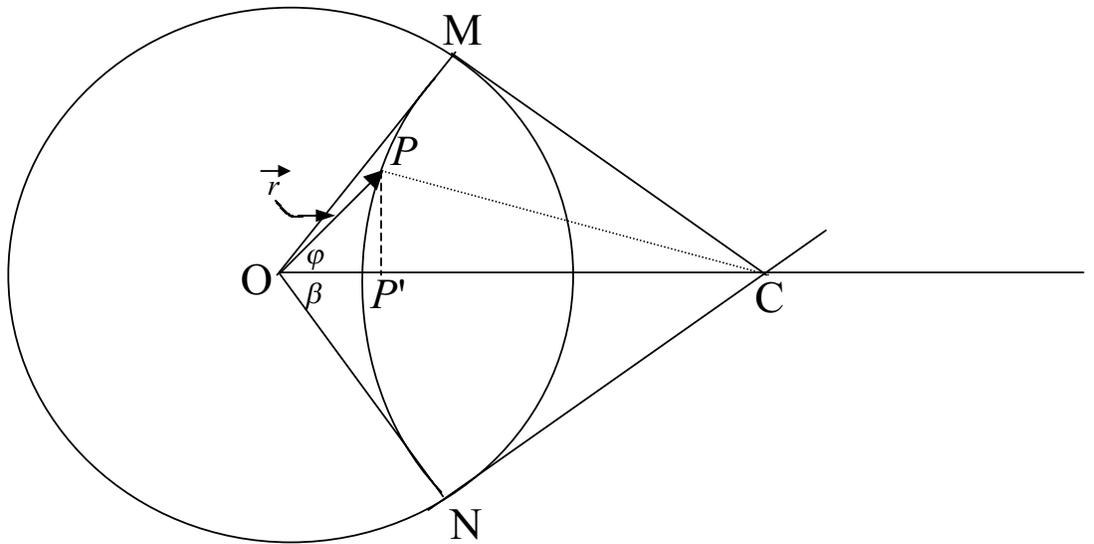,width=0.8\linewidth}
\end{center}
\caption{The outer circle is the intersection of the plane OMN with the
2-sphere boundary of the Poincar\'{e} ball. The geodesic connecting M to N
is a circle whose centre is at C, and it intersects the boundary
orthogonally at points M and N.}
\label{Fig3}
\end{figure}

\begin{figure}[tbp]
\begin{center}
\epsfig{file=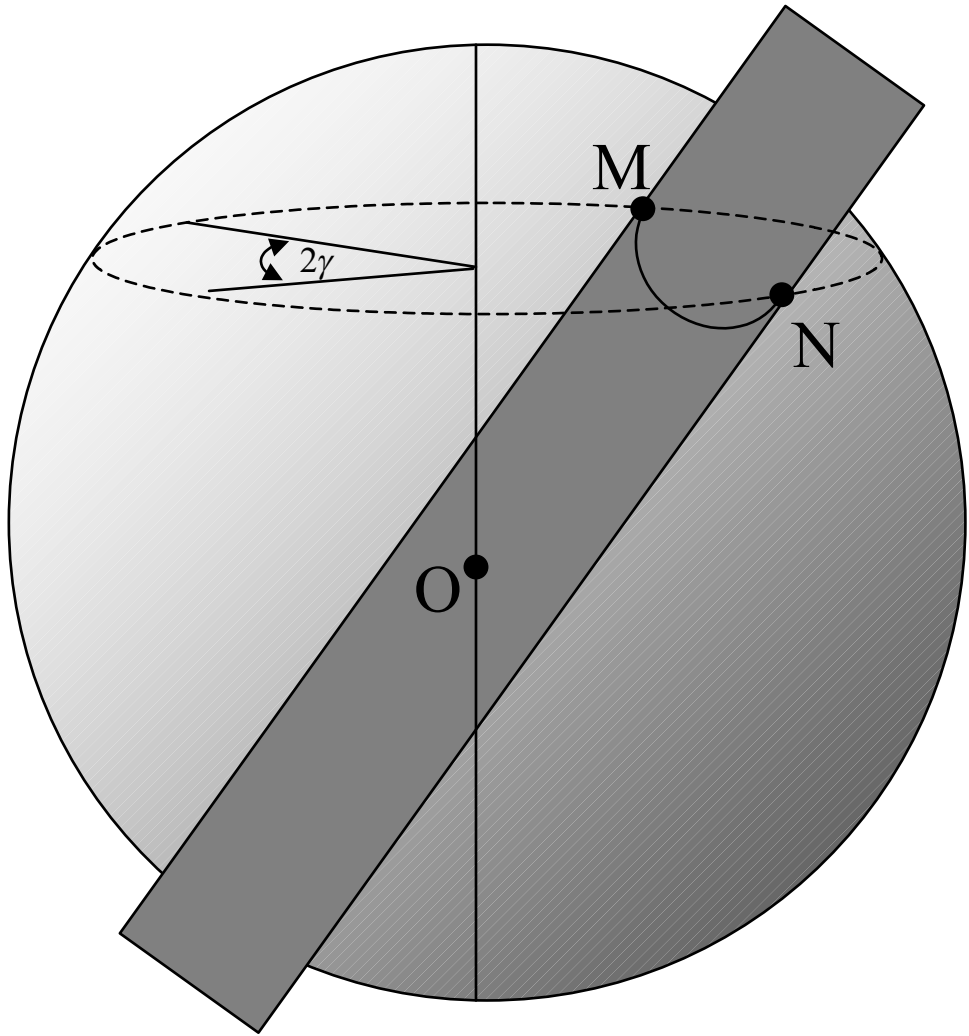,width=0.8\linewidth}
\end{center}
\caption{The intersection of the plane OMN with the Poincar\'{e} ball. A
deficit angle of $2\protect\gamma $ is cut out as shown.}
\label{Fig4}
\end{figure}

To calculate the geodesic length, we parametrize the above geodesics (\ref
{geo1}) and (\ref{geo2}) by 
\begin{equation}
\sinh \chi =\sqrt{\frac{a^{^{\prime ^{2}}}+\varsigma ^{2}}{1-a^{^{\prime
^{2}}}}},\text{ \ }\cos \theta =\cos \theta _{0}\sqrt{\frac{1+\varsigma ^{2}%
}{a^{\prime ^{2}}+\varsigma ^{2}}},\text{ \ }\cos \varphi =b\cos \theta _{0}%
\sqrt{\frac{1+\varsigma ^{2}}{a^{\prime ^{2}}-1+(\varsigma ^{2}+1)\sin
^{2}\theta _{0}}}  \label{parametr}
\end{equation}
where $0\leq \varsigma \leq \infty ,$ and then substitute (\ref{parametr})
into 
\begin{equation}
{\cal L}=\int \sqrt{\stackrel{\cdot }{\chi }^{2}+\sinh ^{2}\chi (\stackrel{%
\cdot }{\theta }^{2}+\stackrel{\cdot }{\varphi }^{2}\sin ^{2}\theta )}%
d\varsigma  \label{L}
\end{equation}
where the over-dot means differentiation with respect to the parameter $%
\varsigma .$ After integration over $\varsigma $, we obtain 
\begin{equation}
{\cal L}=2\ln \{\sinh \chi \sin \theta _{0}\sin \varphi +\sqrt{1+\sinh
^{2}\chi \sin ^{2}\theta _{0}\sin ^{2}\varphi }\}
\end{equation}
for the geodesic length between \ the points $M$ and $N$. In the limit of $%
R\rightarrow 1$ or $\varepsilon \rightarrow 0$ this becomes 
\begin{equation}
{\cal L}=2\ln (\frac{2}{\varepsilon }\sin \theta _{0}\sin \varphi )
\label{len}
\end{equation}

\subsection{Conformal Field Theory Description of the Vortex Structure}

We want to replace the vortex by a geometrical structure in the spacetime.
We have seen in the preceding section that the metric is that of (\ref
{adsmetr1}) with a deficit angle $2\gamma $. Then the constant time
hypersurface in (\ref{ads4met}) is a Poincar\'{e} ball with a wedge cut from
it that runs from $\theta =0$ to $\theta =\pi $. In any plane orthogonal to
the $z$-axis, we have a wedge with angle $2\gamma $ with its boundary edges
identified.

\bigskip

From (\ref{ads4met}) the boundary of AdS$_{4\text{ }}$is ${\bf R\times S}%
^{2} $. Any closed curve on this boundary could be parametrized as $%
(t(\sigma ),\theta (\sigma ),\varphi (\sigma )),$ where the range of $\sigma 
$ is taken $0\leq \sigma \leq 2\pi .$ Now let us take any two points on the
constant time boundary, i.e. on the sphere at the same polar angle $\theta
_{0}$ with radius very close to the radius of boundary of Poincar\'{e} ball
and opposite sign azimuth angles. In this case, where $t$, $\theta $ are
constant, one can take the parameter $\sigma $ to be the same as $\varphi .$
Some of the geodesics passing through the two points which correspond to
different values of $\sigma $ (or $\varphi $) intersect the edge of the
wedge and some other do not.

\bigskip

Following ref. \cite{Bala}, we take coordinate systems $C_{1\text{ }}$ and $%
C_{2}$. In the first coordinate system, $C_{1}$, we take the range of
deficit angle from $\pi -\gamma $ to $\pi +\gamma $, so the range of
variation of polar coordinate $\varphi $ is $0\leq \varphi \leq \pi -\gamma
, $ $\pi +\gamma \leq \varphi \leq 2\pi $. In the second coordinate system $%
C_{2}$, we take the deficit angle to be from $-\gamma $ to $+\gamma $. In
this case the actual value of $\varphi $ changes between $\gamma $ and $2\pi
-\gamma $.

When one increases the value of $\varphi $ from zero to some specific value,
there exists an angle $\varphi _{c}$ for which the logarithmic derivative of
the correlation function has a discontinuity. The reason is as follows. \ It
is a simple matter to see that the geodesics which intersect the
identification in $C_{1\text{ }}$coordinate system do not intersect the
identification in $C_{2\text{ }}$coordinate system and vice-versa. Hence in
evaluating the length of the geodesics between two points which intersect
the wedge of $C_{1},$ it is convenient to transform to $C_{2}$ coordinates.
The effect of this is just to add a constant $\pm \gamma $ to the polar
coordinate $\varphi $ of $C_{1}$ depending to its sign -- $+\gamma $ for
positive $\varphi $ and $-\gamma $ for negative $\varphi $. One can easily
find that if the angle $\varphi $ is less than the critical polar angle ,
namely $\varphi _{c}=\frac{\pi -\gamma }{2}$, then the geodesics do not pass
through the wedge $2\gamma .$ In this case, the length of geodesics should
be computed in $C_{1}$, which is equal to (\ref{len}). On the other hand, if
the angle $\varphi $ is greater than the critical value $\varphi _{c}$, then
the geodesics intersect the edges of the wedge and we must use the
coordinate system $C_{2}$, which the length of geodesic is given by
replacing $\varphi \rightarrow \varphi +\gamma $ in (\ref{len}).

{\bf \bigskip }

Hence there will be a discontinuity in the logarithmic derivative of the
correlation function just in the neighborhood of $\varphi _{c}.$ The amount
of this discontinuity is 
\begin{equation}
K=\frac{1}{<{\cal OO}>}\frac{d<{\cal OO}>}{d\sigma }(\sigma _{c+})-\frac{1}{<%
{\cal OO}>}\frac{d<{\cal OO}>}{d\sigma }(\sigma _{c-})=2\Delta \frac{%
\partial \varphi (\sigma _{c})}{\partial \sigma }\tan \frac{\gamma }{2}
\label{kinkcft}
\end{equation}
where as mentioned above, $\frac{\partial \varphi (\sigma _{c})}{\partial
\sigma }$ is some constant $C_{0}$, which could be set equal to $1$.
Although the correlation function depends on $\theta _{0}$, the magnitude of
the kink is independent of this quantity. We will return to this point later.

\bigskip

\section{The Vortex Mass in AdS$_{4}$}

In this section we compute the mass of the vortex, which is equivalent to
the mass of AdS$_{4}$ with a deficit angle of $\ 2\gamma $. \ 

We choose a two-dimensional manifold ${\cal B}$, which is the boundary of
the Poincar\'{e} ball at $R=l$, or $r=r_{0}\rightarrow \infty $. The
boundary stress tensor is 
\begin{equation}
T^{\mu \nu }=\frac{1}{8\pi }\{\Theta ^{\mu \nu }-\Theta \gamma ^{\mu \nu }+%
\frac{2}{\sqrt{-\gamma }}\frac{\delta I_{ct}}{\delta \gamma _{\mu \nu }}\}
\label{stress}
\end{equation}
where $\gamma ^{\mu \nu \text{ }}$is the induced metric on the boundary of
AdS$_{4}$ ($\partial $AdS$_{4}$), located at $R=l$ with extrinsic curvature $%
\Theta ^{\mu \nu }$. The counterterm action $I_{ct}$ is given by \cite{Mann} 
\begin{equation}
I_{ct}=\frac{1}{4\pi l}\int_{\partial AdS_{4}}d^{3}x\sqrt{-\gamma }(1+\frac{%
l^{2}}{4}{\cal R})  \label{Ict}
\end{equation}
which, when added to the usual Einstein-Hilbert action of AdS-gravity,
removes its divergences at $r\rightarrow \infty $.

The mass is given by \cite{Brown}

\begin{equation}
M=\frac{1}{8\pi }\int_{{\cal B}}d^{2}x\sqrt{\sigma }T_{\mu \nu }u^{\mu }\xi
^{\nu }  \label{mass}
\end{equation}
where $u^{\mu }$ is the timelike normal unit vector to the boundary ${\cal B}
$ with the metric $\sigma ^{ab}$, and defines the local arrow of time on the
boundary of AdS$_{4\text{ }}$. $\xi ^{\nu }$ is the time-like Killing vector
of AdS$_{4}$.

The deficit angle in the AdS$_{4\text{ }}$space yields a singular structure
in the induced Ricci scalar of the boundary ${\cal B}$. These conical
singularity structures are due to the identification of the edges of the
wedge at the points $\theta =0$ and $\theta =\pi $ on the ${\cal B}$. To
handle these singularities,\thinspace\ we replace the boundary ${\cal B}$
with a sequence of \ regular manifolds \cite{Mann2,Solo}. The first and
second regular manifolds $\widetilde{{\cal M}_{\gamma }^{(0)}}$, $\widetilde{%
{\cal M}_{\gamma }^{(\pi )}}$ are regulated manifolds corresponding to the $%
{\cal M}_{\gamma }^{(0)}$, ${\cal M}_{\gamma }^{(\pi )}$ which are two
dimensional spaces with the topology of a cone in the neighborhood of the
poles of the boundary ${\cal B}$. The third manifold is ${\cal B}/{\cal (}%
{\cal M}_{\gamma }^{(0)}\otimes $ ${\cal M}_{\gamma }^{(\pi )})$, which can
be smoothly matched to the other two. On the manifolds ${\cal M}_{\gamma
}^{(0)}$, ${\cal M}_{\gamma }^{(\pi )}$, \ the metric is, 
\begin{equation}
r_{0}^{2}(d\theta ^{2}+\theta ^{2}d\varphi ^{2})  \label{conmetr}
\end{equation}
where the polar coordinate $\varphi $ is restricted to the interval $%
[0,2(\pi -\gamma )].$ We take the following form of the metric for the
regular manifolds \ $\widetilde{{\cal M}_{\gamma }^{(0)}}$, $\widetilde{%
{\cal M}_{\gamma }^{(\pi )}}$,\ 
\begin{equation}
r_{0}^{2}\{(1-\frac{\gamma }{\pi })^{2}+\frac{\gamma }{\pi }(\frac{\gamma }{%
\pi }-2)[\frac{\partial f(\theta ,a)}{\partial \theta }]^{2}\}d\theta
^{2}+r_{0}^{2}\theta ^{2}d\varphi ^{2}  \label{conregmetr}
\end{equation}

\ \ \ \ \ In the above equation, $f(\theta ,a)$ is a function which is
introduced to smooth off the tips of the cones,\ and is subject to the
following conditions , 
\begin{equation}
\lim_{a\rightarrow 0}f(\theta ,a)=\theta ,\text{ }\lim_{a\rightarrow 0}\text{
}\frac{\partial f(\theta ,a)}{\partial \theta }=0\text{\ }
\end{equation}
where $a$ is the regulator parameter. By choosing an appropriate function
form of $\ f(\theta ,a)$ \cite{Solo}, it is easy to show that the Ricci
scalar ${\cal R}_{\gamma }$ of the boundary ${\cal B}$ with conical
singularities at $\theta =0$ \ and $\theta =\pi $ is \ 
\begin{equation}
{\cal R}_{\gamma }={\cal R}+\frac{2\gamma }{\pi -\gamma }\frac{1}{r_{0}^{2}}%
\{\delta (\theta )+\delta (\pi -\theta )\}  \label{conricci}
\end{equation}
In the above relation, ${\cal R}$ is the Ricci scalar of the manifold ${\cal %
B}/{\cal (}{\cal M}_{\gamma }^{(0)}\otimes $ ${\cal M}_{\gamma }^{(\pi )})$
and the Dirac delta function is subject to the following normalization, 
\begin{equation}
\int \theta \delta (\theta )d\theta =1  \label{normal}
\end{equation}

Putting these together, (\ref{mass}) gives for the mass of AdS$_{4}$,
containing a wedge , 
\begin{equation}
M=\frac{\gamma }{2\pi }r_{0}  \label{totmass}
\end{equation}
in agreement with (\ref{massdef}) in the limit that the thickness $\rho _{0}$
of the string is negligible.

\bigskip

\section{Conclusion}

\bigskip

We have solved the Nielsen-Olesen equations in an AdS$_{4}$ background, and
found that the Higgs and gauge fields are axially symmetric, with non-zero
winding number. Our solution in the limit of \ large $l$ (small cosmological
constant) reduces to the well known flat-space \cite{NO}, and has
well-defined characteristic lengths $\lambda $ and $\beta $ outside of which
the fields exponentially approach their asymptotic values. The solution (to
leading order in the gravitational coupling) induces a deficit angle in AdS$%
_{4}$. We find these results compelling enough to interpret our solution as
a vortex.

The holographic detection of a Nielsen-Olesen vortex is now obtained by
comparing eqs. (\ref{kinkcft}) and (\ref{totmass}). Although the total mass
of the vortex is infinite, the mass per unit length is finite, and we can
relate this to the kink (\ref{kinkcft}) in the correlation function computed
above. Since the kink depends only on the deficit angle $\gamma $, therefore
for a fixed $\gamma ,$ the kink is constant. On the other hand mass density
also depends only on $\gamma $, and therefore one could interpret the value
of the kink as the mass per unit length. From (\ref{totmass}), the mass per
unit length is equal to $\mu =\frac{M}{2r_{0}}=\frac{\gamma }{4\pi }$, {\bf %
\ }and we obtain 
\begin{equation}
\frac{M}{2r_{0}}=\frac{1}{2\pi }\arctan \left( \frac{K}{2C_{0}\Delta }\right)
\label{masskink}
\end{equation}
for the relationship between the vortex mass density and the kink in the CFT
correlation function. \ 

The winding number of the vortex also has a holographic description. \ The
field $\Phi $ in (\ref{XPomegadef}) has the same phase as its limiting field
at the boundary of AdS$_{4}$ . Consequently the winding number of the vortex
is given by the line-integral of the corresponding one-point function about
any loop which encloses $\theta =0$ on the $(2+1)$-dimensional boundary.

Of course another holographic effect of the vortex is to induce conical
singularities in the boundary at $\theta =0,\pi $\ . The holographic
interpretation is that the boundary CFT is formulated on a space that has a
conical deficit. \ However note that the presence of such singularities need
not imply the presence of a kink in the correlation function. \ It is this
kink -- present at any latitude of the boundary spacetime -- that signals
the presence of a vortex.

Our results are commensurate with those of ref. \cite{Bala}: \ non-local
properties of the two point correlation function in the dual conformal field
theory (the kink, in this case) probe the interior of the bulk geometry. \
We have considered only the two point function of the operator on the
boundary which is dual to the scalar field $\Phi $\ in the Lagrangian (\ref
{Lag}). \ As we have seen, this correlation function is related to the mass
density of the vortex. An understanding of the role of the other two, three
and four points correlation functions remains to be carried out, along with
a more detailed study of the vortex itself, including obtaining a
holographic description of its interior structure, and of threading it
through a black hole. Work on these problems is in progress.

\bigskip

{\Large Acknowledgments}

This work was supported by the Natural Sciences and Engineering Research
Council of Canada.

\end{document}